\documentclass[letterpaper,prl,floatfix,showpacs,twocolumn]{revtex4-1}%-1
\usepackage{amssymb,amsmath,graphicx,xspace,overpic,ifthen,verbatim, subfigure, appendix}
\pdfoutput=1

\DeclareMathOperator{\Tr}{Tr}

\newcommand{\dd}{\,\mathrm{d}}
\newcommand{\arctanh}{\,\mathrm{arctanh}}

\newcommand{\MI}{\ensuremath{\mathcal{I}}}
\newcommand{\Sys}{\ensuremath{\mathcal{S}}}
\newcommand{\Env}{\ensuremath{\mathcal{E}}}
\newcommand{\Frag}{\ensuremath{\mathcal{F}}}

\newcommand{\pbwidthfactor}{0.96}

\begin{document}
%%%%%%%%%%%%%%%%%%%%%%%%%%%%%

\def\FCW{0.98\columnwidth}
\def\HPW{0.48\textwidth}
\def\TQPW{0.73\textwidth}
\def\FPW{0.98\textwidth}

%\graphicspath{{ShortFigs/}}

%%%%%%%%%%%%%%%%%%%%%%%%%%%%%
%      Title/authors        %
%%%%%%%%%%%%%%%%%%%%%%%%%%%%%

\title{Quantum Darwinism in an Everyday Environment: Huge Redundancy in Scattered Photons}
\date{\today}
\author{C.~Jess~Riedel$^{1}$ and Wojciech~H.~Zurek$^1$}\
\affiliation{$^1$Theory Division, LANL, Los Alamos, New Mexico 87545, USA}

%%%%%%%%%%%%%%%%%%%%%%%%%%%%%
%         Abstract          %
%%%%%%%%%%%%%%%%%%%%%%%%%%%%%

\begin{abstract}
We study quantum Darwinism---the redundant recording of information about the preferred states of a decohering system by its environment---for an object illuminated by a blackbody. In the cases of point-source and isotropic illumination, we calculate the quantum mutual information between the object and its photon environment.  We demonstrate that this realistic model exhibits fast and extensive proliferation of information about the object into the environment and results in redundancies orders of magnitude larger than the exactly soluble models considered to date.
\end{abstract}


\maketitle

%%%%%%%%%%%%%%%%%%%%%%%%%%%%%
%      Introduction         %
%%%%%%%%%%%%%%%%%%%%%%%%%%%%%

The theory of decoherence \cite{SchlosshauerText,*JoosText,*Zurek2003} is the foundation of the modern understanding of the quantum-classical transition.  However, the standard analysis largely ignores the environment by tracing over it when, in fact, the environment plays a crucial role in how real observers find out about the world.  Classically, properties of systems are objective in that they can be independently measured and agreed upon by \emph{arbitrarily} many observers without disturbing the system itself.  This can arise in a purely quantum universe when many copies of information about a system's properties are imprinted onto the environment.  Under the condition of effective decoherence, this information can only describe the pointer states of the system, \emph{not} superpositions thereof \cite{Zurek2009}.  In this sense, pointer states are distinguished not only for forming the stable basis in which the density matrix of the system diagonalizes but also for being redundantly copied into the environment. Quantum Darwinism \cite{Zurek2000, *Ollivier2004} is a framework for analyzing the flow of information about these ``fittest'' states, helping to elucidate the origin of classical objectivity.

Quantum Darwinism has been investigated for a spin\nobreakdash-$\frac{1}{2}$ particle monitored by a pure \cite{Blume-Kohout2005} and mixed \cite{Zwolak2009, Zwolak2009a} bath of spins and a harmonic oscillator monitored by a pure bath of oscillators \cite{Blume-Kohout2008, Paz2009}.  These studies support the intuition that redundant proliferation of pointer-state information should be common in decohering systems, but the models are abstract and limited in size by the feasibility of numerical calculations.  Therefore, they do not allow one to estimate the redundancies expected in physically realistic systems.  In this Letter we shed light on this question by considering an object illuminated by a black body.  

When an object in a macroscopic superposition is exposed to radiation, scattering photons will quickly reduce its pure, nonlocal state to a mixture of localized alternatives via collisional decoherence \cite{Joos1985}.  (See also \cite{Diosi1995,Gallis1990,Hornberger2003,*Hornberger2006} for refinements and corrections.) The fantastic rate of collisional decoherence has been confirmed experimentally \cite{Kokorowski2001,Uys2005}. Here we investigate this process as an example of quantum Darwinism.
%\cite{Arndt1999}

%%%%%%%%%%%%%%%%%%%%%%%%%%%%%
%      Intro Diagram        %
%%%%%%%%%%%%%%%%%%%%%%%%%%%%%

\begin{figure} [bp]
  \centering 
  \includegraphics[width=\pbwidthfactor\columnwidth]{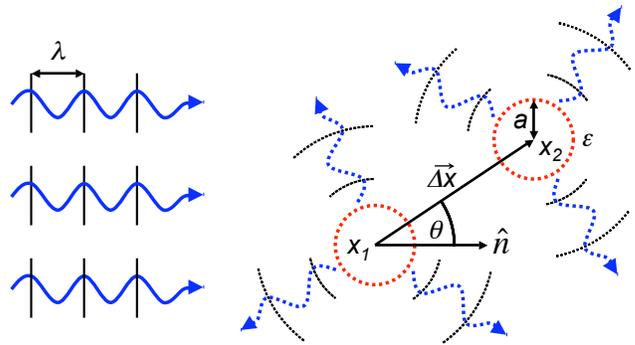}
  \caption{A dielecric sphere of radius $a$ and permittivity $\epsilon$ is initially in a superposition with separation $\Delta x = |x_1-x_2|$.  The object is subjected to plane-wave radiation with thermally distributed wavelength $\lambda$ and propagating in a direction $\hat{n}$ that makes an angle $\theta$ with the vector $\vec{\Delta x}$.}
  \label{intro_diagram}
\end{figure}

%%%%%%%%%%%%%%%%%%%%%%%%%%%%%%%%%

Following Joos and Zeh \cite{Joos1985}, we consider a dielectric sphere of radius $a$ and relative permittivity $\epsilon$ initially in a superposition, so that $|\psi (\vec{x})|^2 \approx [\delta(\vec{x}-\vec{x}_1)+\delta(\vec{x}-\vec{x}_2)]/2$ for some $\Delta x = |\vec{x}_1-\vec{x}_2|$ (see Fig.\ \ref{intro_diagram}).  We ignore the self-Hamiltonian of the object so that its effective Hilbert space $\Sys$ is spanned by the two position eigenstates.  The sphere is illuminated from a single direction by radiation from a point-source black body at a temperature $T$. 

Observers typically access a small part of the environment (in this case, the photons that enter one's eye), so we will estimate how much information about the object is available in a subset of the environmental photons.  We assume our environment consists of a large but fixed number $N$ of photons: $\Env = \bigotimes_{n = 1}^N \Env_i$, where $\Env_i$ is the Hilbert space of a single photon in a box of volume $V$.  We then define $\Frag_f = \bigotimes_{n = 1}^{fN} \Env_i$ to be the \emph{fragment} corresponding to some fraction $f$ of the environment composed of $f N$ photons.  Since the photons have identical initial conditions and interactions, the choice of photons with which to construct the fragment is unimportant.  To get our final results, we will take $V$ and $N$ to infinity while holding the photon density $N/V$ constant.

The primary quantity investigated will be the quantum mutual information $\MI_{\Sys : \Frag} = H_{\Sys} + H_{\Frag} - H_{\Sys, \Frag}$ between the system $\Sys$ and fragment $\Frag$, where $H$ denotes the von Neumann entropy.  From this we will calculate the \emph{redundancy} $R_{\delta}$, which is the number of distinct fragments in the environment that supply, up to an \emph{information deficit} $\delta$, the classical information about the state of the system.  More precisely, $R_{\delta} = 1/f_\delta$, where $f_\delta$ is the smallest fragment such that $\MI_{\Sys : \Frag_{f_\delta}} = (1-\delta)H_{\Sys}$.  (Only very large fragments $f \geq 0.5$ will have complete classical information about the object \cite{Blume-Kohout2005}.)

%%%%%%%%%%%%%%%%%%%%%%%%%%%%%
%       Setup Quantum       %
%%%%%%%%%%%%%%%%%%%%%%%%%%%%%

The sphere and the photons in the environment are assumed to be initially unentangled: $\rho^0 = \rho_{\Sys}^0~\otimes~\rho_{e}^0~\otimes~\cdots~\otimes~\rho_{e}^0$, where $\rho_{\Sys}$ and $\rho_{e}$ are the density matrices of the system and of a single photon, respectively, and a superscript ``$0$'' denotes prescattering states.  The photon momenta are distributed according to $\rho_{e}^0 = \int_0^{\infty} \dd k \, p(k) k^2|\vec{k}\rangle\langle\vec{k}|$ for $p(k) \propto  k^2/[\exp(k c / k_B T)-1]$ and $\hat{n} = \hat{k_i}$ a unique direction.  

The decoherence of the superposition is governed by
\begin{equation}
\label{position_decoh}
|\langle \vec{x_1} | \rho_\Sys | \vec{x_2}\rangle|^2 = \gamma^N |\langle \vec{x_1} | \rho_{\Sys}^0 | \vec{x_2}\rangle|^2 ,
\end{equation}
where 
\begin{equation}
\gamma  \equiv \left| \Tr \left[ S_{\vec{x}_1} \rho_{e}^0 S_{\vec{x}_2}^\dagger \right] \right|^2 \ 
\end{equation}
and $S_{\vec{x}_p}$ is the scattering matrix acting on the single photon state when the particle is located at $\vec{x}_p$. Because $\gamma$ controls the suppression of the off-diagonal terms of the object's density matrix in the position basis, $\gamma$ and $\Gamma \equiv \gamma^N$ are the \emph{decoherence factors} attributable to a single scattering photon and the environment as a whole, respectively.  The two-dimensional $\rho_\Sys$ can be diagonalized and its entropy is
\begin{align}
\label{H_S_open}
H_{\Sys} &= \ln 2 - \sum_{n=1}^{\infty} \frac{\Gamma^{n}}{2n(2n-1)} \\
\label{H_S_closed}
&= \ln 2 - \sqrt{\Gamma} \arctanh{\sqrt{\Gamma}} - \ln \sqrt{1-\Gamma},
\end{align}

We use the classical cross section of a dielectric sphere \cite{JacksonEMcrosssection} in the dipole approximation ($\lambda \gg a$) and assume the photons are not sufficiently energetic to resolve the superposition individually ($\lambda \gg \Delta x$).  We further assume that the object is heavy enough to have negligible recoil and that photon energy is conserved.  Under these conditions, the key matrix element (which coincides with $\gamma$ in the case of \emph{monochromatic} radiation) is
\begin{align}
\begin{split}
|\langle \vec{k}(\lambda)|{S_{\vec{x}_1}}^\dagger S_{\vec{x}_2} |\vec{k}(\lambda)\rangle|^2 &= \\  1-\frac{1}{V} \frac{256 \, \pi^7}{15} &(3+11 \cos^2 \theta) \frac{\tilde{a}^6 \Delta x^2 t c}{\lambda^6} 
\end{split}
\end{align}
to leading order in $1/V$. Above, $\tilde{a} \equiv a [(\epsilon-1)/(\epsilon-2)]^{1/3}$ is the effective radius of the object and $t$ is the elapsed time.  The states $|\vec{k}(\lambda)\rangle$ are photon momentum eigenstates with wavelength $\lambda$ making an angle $\theta$ with the separation vector $\vec {\Delta x}$. 

For increasing $V$, photon momentum eigenstates become diffuse so individual photons decohere the state less and less (i.e. $\gamma \rightarrow 1$).  This is balanced, of course, by an increasing number of photons in the box, which will lead to a finite decoherence factor for the whole environment, $\Gamma = \gamma^N$. In the $V \to \infty$ limit we use $e = \lim_{q \to \infty} (1 + 1/q)^q$ to get $\Gamma = \exp (-t/\tau_D)$, where \footnote{The decoherence rate does not increase for arbitrarily large $\Delta x$.  Rather, this expression is only valid in the $\Delta x \ll \lambda$ limit we are considering here.  For $\Delta x \gg \lambda$, the decoherence rate saturates
\begin{align*}
\frac{1}{\tilde{\tau}_D} = \tilde{C}_\Gamma \frac{I \tilde{a}^6 k_B^3 T^3}{c^4 \hbar^4}
\end{align*}
with $\tilde{C}_\Gamma = 57600 \, \zeta(7)/\pi^3 \approx 1873$, in agreement with \cite{Gallis1990}.  In the intermediate region, $\Delta x \sim \lambda$, $\tau_D$ has a complicated dependence on both $\Delta x$ and $\theta$.  The results of the present work are valid for \emph{all} $\Delta x$ so long as the correct $\tau_D$ is used.}
\begin{equation}
\label{pure_decoh_time}
\frac{1}{\tau_D} = C_\Gamma (3+11 \cos^2 \theta) \frac{ I \tilde{a}^6 \Delta x^2 k_B^5 T^5}{c^6 \hbar^6}.
\end{equation}
and $C_\Gamma = 161280	\, \zeta(9)/\pi^3 \approx 5210$ is a numerical constant. We have replaced the photon density $N/V$ with the more physical \emph{irradiance} $I$ (radiative power per unit area).  Given Eq.\ (\ref{position_decoh}), we identify $\tau_D$ as the \emph{decoherence time}.  Although the rate of decoherence (and, as we shall see, the redundancy) depends on the angle of illumination $\theta$, decoherence is usually so rapid that it hardly matters.

To get the mutual information, we can avoid calculating $H_{\Sys \Frag}$ by using the identity [Eq. (8) of \cite{Zwolak2009a}]
\begin{align}
\MI_{\Sys : \Frag} = \left[H_{\Frag} - H_{\Frag}^0 \right] + \left[ H_{\Sys d \Env} - H_{\Sys d \Env / \Frag} \right]
\end{align}
where $H_{\Sys d \Env} = H_{\Sys}$ is the entropy of the system as decohered by the entire environment $\Env$ and $H_{\Sys d \Env / \Frag}$ is the entropy of the system if it were decohered by only $\Env/\Frag$.  We obtain $H_{\Sys d \Env / \Frag}$ from $H_{\Sys}$, Eq. \eqref{H_S_open}, by making the replacement $\Gamma \to \Gamma^{1-f}$. Despite the mixedness of the environment, it is possible to diagonalize the post-scattering state $\rho_{\Frag}$ to get $H_\Frag$ because of the special form of our model; the photons are of mixed energy but are in directional eigenstates, while the elastic scattering conserves energy but mixes photon direction.  This allows us to write
\begin{align}
\rho_{\Frag}  &= \int \dd \chi_\Frag \, p(\chi_\Frag) \, |\chi_\Frag \rangle \langle \chi_\Frag| \otimes \rho_{\hat{\Frag}}^\chi \, \, , \\
\rho_{\hat{\Frag}}^\chi &= \frac{1}{2} \left[ \bigotimes_{i=1}^{f N} S_{\vec{x}_1}^{k_i} |\hat{n}\rangle\langle \hat{n}| {S_{\vec{x}_1}^{k_i}}^\dagger + \bigotimes_{i=1}^{f N} S_{\vec{x}_2}^{k_i} |\hat{n}\rangle\langle \hat{n}| {S_{\vec{x}_2}^{k_i}}^\dagger \right],
\end{align}
where we have broken the momentum eigenstates into a tensor product $|\vec{k}_i\rangle = |k_i\rangle|\hat{n}\rangle/k_i$ of magnitude and directional eigenstates. Above, $\chi_\Frag = (k_1,\ldots, k_{f N})$ is the vector of the magnitudes of the photon momenta of \Frag, $p(\chi_\Frag) = \prod_{i=1}^{f N} p(k_i) $ is the momentum probability distribution, and $|\chi_\Frag\rangle\langle\chi_\Frag| = \bigotimes_{i=1}^{f N} |k_i\rangle\langle k_i|$.  $S_{\vec{x}_p}^{k_i}$ is defined by
\begin{align}
S_{\vec{x}_p} |\vec{k}_i \rangle = S_{\vec{x}_p} |k_i\rangle |\hat{n}\rangle /k = |k_i\rangle S_{\vec{x}_p}^{k_i} |\hat{n}\rangle /k \, .
\end{align}

We then have
\begin{align}
\label{H_F}
H_{\Frag} &=  f N H_e^0 + \int \dd \chi_\Frag \, p(\chi_\Frag) \, H_{\hat{\Frag}}^{\chi} \, ,
\end{align}
where $H_{\hat{\Frag}}^{\chi}$ is the entropy of $\rho_{\hat{\Frag}}^{\chi}$ and $H_e^0 = H_\Frag^0 / f N$ is the initial entropy of a single thermal photon (which diverges since the photon Hilbert space is infinite dimensional).  Although the conditional state $\rho_{\hat{\Frag}}^{\chi}$ lives in an infinite-dimensional vector space, it has only two nonzero eigenvalues,
\begin{align}
\lambda^\chi_{\hat{\Frag}}  &= \frac{1}{2} \pm  \frac{1}{2} \prod_{i = 1}^{f N} \left| \langle\hat{n}|{S_{\vec{x}_1}^{k_i}}^\dagger S_{\vec{x}_2}^{k_i} |\hat{n}\rangle \right| .
\end{align}
After plugging these into the formula for entropy, we can perform the integral in Eq.\ \eqref{H_F}.  The divergent pieces cancel in the mutual information and we are left with
\begin{align}
\label{mixed_MI}
\MI_{\Sys : \Frag_f} &= \ln 2 + \sum_{n=1}^{\infty} \frac{\Gamma^{(1-f)n} - \Gamma^{f n} -\Gamma^{n}}{2n(2n-1)} ,
\end{align}

\begin{figure} [tb!]
  \label{QMI_plots}
  \centering 
  \subfigure[Point-source illumination]{
    \label{mixed_QMI_plot}
    \includegraphics[width=\pbwidthfactor\columnwidth]{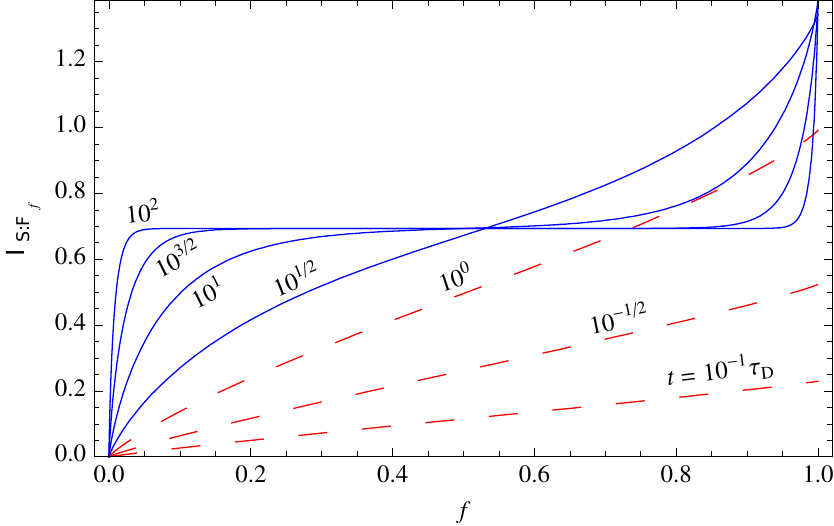}
  }
  \subfigure[Isotropic illumination]{
    \label{isotropic_QMI_plot}
    \includegraphics[width=\pbwidthfactor\columnwidth]{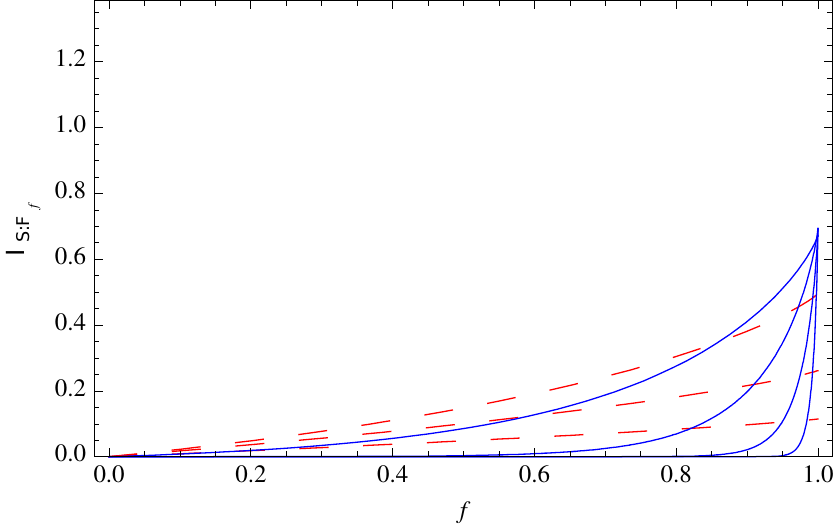}
  }
  \caption{The quantum mutual information $\MI_{\Sys : \Frag_f}$ versus fragment size $f$ at different elapsed times for an object illuminated by (a) point-source blackbody radiation, Eq. \eqref{mixed_MI}, and (b) isotropic blackbody radiation, Eq. \eqref{isotropic_MI}. \\ 
(a) For point-source illumination, individual curves are labeled by the time $t$ in units of the characteristic time $\tau_D$, Eq.\ \eqref{pure_decoh_time}. For $t \le\tau_D$ (red dashed lines), the information about the system available in the environment is low.  The linearity in $f$ means each piece of the environment contains new, independent information.  For $t>\tau_D$ (blue solid lines), the plateau shape of the curve indicates redundancy; the first few pieces of the environment give a large amount of information, but additional pieces just confirm what is already known. On the plateau, the mutual information approaches it's maximum classical value, $\MI_{\Sys : \Frag_f} = 1$ bit $= \ln 2$ nats $\approx 0.69$ nats. The remaining information (i.e., above the plateau) is highly \emph{encoded} in the global state, in the sense that it can only read by capturing almost all of $\Env$. \\
(b) For isotropic illumination, the same time slicing is used but there is greatly decreased mutual information because the directional photon states are already ``full'' and cannot store more information about the state of the object.  Zero redundant copies are produced and the mutual information approaches $0$ as $t \to \infty$ for all $f < 1$.}
\end{figure}

The mutual information is plotted in Fig.\ \ref{mixed_QMI_plot} as a function of $f$ for different values of $t$.  This is a \emph{partial information plot} \cite{Blume-Kohout2005, Blume-Kohout2008, Zurek2009}.  It charts how much information about the object is available to an observer depending on the size of the fragment captured.  For times $t \gg \tau_D$, the mutual information has a distinctive plateau that indicates redundancy.  

%%%%%%%%%%%%%%%%%%%%%%%%%%%%%
%    Quantum PIP Figure     %
%%%%%%%%%%%%%%%%%%%%%%%%%%%%%

%%%%%
%%%%%
%%%%%

The summations in \eqref{mixed_MI} can be written in a closed form analogous to Eq. \eqref{H_S_closed}, but the power series is more useful for calculating the redundancy.  For large times, $\Gamma = \exp(-t/\tau_D)$ is exponentially small and the sum is dominated by the lowest power of $\Gamma$.  If $f<1/2$, then $f < (1-f) < 1$ and
\begin{align}
\label{mixed_MI_approx}
\MI_{\Sys : \Frag_f} &\approx \ln 2 - \frac{1}{2} \Gamma^{f} . %\quad \qquad (t \gg \tau_D, f<\frac{1}{2}).
\end{align}
This allows us to estimate the redundancy (for $\delta < 0.5$) in the limit $t \gg \tau_D$:
\begin{align}
\label{pure_redundancy}
R_\delta &= \frac{1}{\ln [(2\, \delta \ln 2)^{-1}]} \frac{t}{ \tau_D} .%\\
%&= \frac{t}{ \tau_R}.
\end{align} 
%where we have defined the redundancy time $\tau_R = \tau_D  \ln (1/ 2\, \delta \ln 2) $.

The key points are these: First, the redundancy depends only weakly (logarithmically) on the information deficit $\delta$, which is consistent with previous results \cite{Blume-Kohout2006, Zwolak2009a}.  Second, the redundancy increases linearly with time at a rate given by the inverse of the decoherence time.  This is intuitive because (1) photons scatter off the object at a constant rate and (2) it is precisely the dependence of photon out states on the position of the object (roughly corresponding to a record) that causes decoherence.  Since even very tiny objects have extremely short decoherence times \cite{SchlosshauerText, Joos1985}, the redundancy quickly becomes enormous.

It should be noted that there will not be strong redundancy if the object is illuminated uniformly from all directions---despite the fact that the rate of decoherence is simply given by averaging Eq.\ \eqref{pure_decoh_time} over the solid angle.  For isotropic illumination, $H_{\Frag} = H_{\Frag}^0$ and
\begin{align}
\MI_{\Sys : \Frag_f} &= H_{\Sys d \Env} - H_{\Sys d \Env / \Frag} \\
\label{isotropic_MI}
&= \sum_{n=1}^{\infty} \frac{\Gamma^{(1-f)n} -\Gamma^{n}}{2n(2n-1)} ,
\end{align}
which, for large times, vanishes for all proper fragments.  This is plotted in Fig.\ \ref{isotropic_QMI_plot}, which shows that the mutual information barely rises from zero before fading away, never yielding a single redundant copy.  This behavior is due to the fact that the component of the environment in which information about the object is stored---the photon directional states---is initially fully mixed and so cannot hold any new information.  

However, this situation is very unnatural since the directional photon states must be perfectly mixed.  In physical situations (e.g., objects lit by light bulbs, the Sun, or ambient light), we expect illumination to be nonuniform and the initial mixedness of the photon environment to decrease the redundancy by only a factor of order unity, in accordance with detailed calculations made of spin\nobreakdash-$\frac{1}{2}$ systems \cite{Zwolak2009a}.  Future research could explore the precise dependence of the redundancy on the degree of directional mixedness as well as investigate the spatial distribution of the redundant information.

%%%%%%%%%%%%%%%%%%%%%%%%%%%%%
%       Conclusion          %
%%%%%%%%%%%%%%%%%%%%%%%%%%%%%

We have shown in this Letter that collisional decoherence, a ubiquitous phenomenon in everyday life, leads to the proliferation of information about objects into the environment at a rate linear in time and (for most systems) on an extremely short time scale.  Indeed, after being illuminated by the Sun for just $1\ \mu$s, a grain of dust $1\ \mu$m across will have its location imprinted about $100\times 10^6$  times in the scattered photons.  Such extensive proliferation allows multiple observers to independently determine an object's position by monitoring the environment, such that the object has an objective, classical location.  The redundancy seen in the photon scattering system is much larger than the abstract examples of quantum Darwinism previously studied because (a) the photon environment, like most real decohering environments, is essentially infinite and (b) the photons that have scattered from the system keep records of its location forever. 

We thank Michael Zwolak and Haitao Quan for helpful discussion. One of us (WHZ) is especially grateful to Charles Bennett, whose penetrating discussion of information propagation and records in (photon and other) environments \cite{Bennett2006} provides a useful setting for the study of quantum Darwinism. This research is supported by the U.S. Department of Energy through the LANL/LDRD program.

%%%%%%%%%%%%%%%%%%%%%%%%%%%%%
%       Bibliography        %
%%%%%%%%%%%%%%%%%%%%%%%%%%%%%
\bibliographystyle{apsrev4-1}%4-1
\bibliography{riedelbib}

%Merlin.mbs v4.21 2009-07-09.
\begin{thebibliography}{10}%
\makeatletter
\providecommand \@ifxundefined [1]{%
 \ifx #1\undefined \expandafter \@firstoftwo
 \else \expandafter \@secondoftwo
\fi
}%
\providecommand \@ifnum [1]{%
 \ifnum #1\expandafter \@firstoftwo
 \else \expandafter \@secondoftwo
\fi
}%
\providecommand \enquote [1]{``#1''}%
\providecommand \bibnamefont  [1]{#1}%
\providecommand \bibfnamefont [1]{#1}%
\providecommand \citenamefont [1]{#1}%
\providecommand\href[0]{\@sanitize\@href}%
\providecommand\@href[1]{\endgroup\@@startlink{#1}\endgroup\@@href}%
\providecommand\@@href[1]{#1\@@endlink}%
\providecommand \@sanitize [0]{\begingroup\catcode`\&12\catcode`\#12\relax}%
\@ifxundefined \pdfoutput {\@firstoftwo}{%
 \@ifnum{\z@=\pdfoutput}{\@firstoftwo}{\@secondoftwo}%
}{%
 \providecommand\@@startlink[1]{\leavevmode\special{html:<a href="#1">}}%
 \providecommand\@@endlink[0]{\special{html:</a>}}%
}{%
 \providecommand\@@startlink[1]{%
  \leavevmode
  \pdfstartlink
   attr{/Border[0 0 1 ]/H/I/C[0 1 1]}%
   user{/Subtype/Link/A<</Type/Action/S/URI/URI(#1)>>}%
  \relax
 }%
 \providecommand\@@endlink[0]{\pdfendlink}%
}%
\providecommand \url  [0]{\begingroup\@sanitize \@url }%
\providecommand \@url [1]{\endgroup\@href {#1}{\urlprefix}}%
\providecommand \urlprefix [0]{URL }%
\providecommand \Eprint[0]{\href }%
\@ifxundefined \urlstyle {%
  \providecommand \doi [1]{doi:\discretionary{}{}{}#1}%
}{%
  \providecommand \doi [0]{doi:\discretionary{}{}{}\begingroup
  \urlstyle{rm}\Url }%
}%
\providecommand \doibase [0]{http://dx.doi.org/}%
\providecommand \Doi[1]{\href{\doibase#1}}%
\providecommand \bibAnnote [3]{%
  \BibitemShut{#1}%
  \begin{quotation}\noindent
    \textsc{Key:}\ #2\\\textsc{Annotation:}\ #3%
  \end{quotation}%
}%
\providecommand \bibAnnoteFile [2]{%
  \IfFileExists{#2}{\bibAnnote {#1} {#2} {\input{#2}}}{}%
}%
\providecommand \typeout [0]{\immediate \write \m@ne }%
\providecommand \selectlanguage [0]{\@gobble}%
\providecommand \bibinfo [0]{\@secondoftwo}%
\providecommand \bibfield [0]{\@secondoftwo}%
\providecommand \translation [1]{[#1]}%
\providecommand \BibitemOpen[0]{}%
\providecommand \bibitemStop [0]{}%
\providecommand \bibitemNoStop [0]{.\EOS\space}%
\providecommand \EOS [0]{\spacefactor3000\relax}%
\providecommand \BibitemShut [1]{\csname bibitem#1\endcsname}%
%</preamble>
\bibitem{SchlosshauerText}%
  \BibitemOpen
  \bibfield{author}{%
  \bibinfo {author} {\bibfnamefont{M.}~\bibnamefont{Schlosshauer}},\ }%
  \emph{\bibinfo {title} {Decoherence and the Quantum-to-Classical
  Transition}}\ (\bibinfo {publisher} {Springer-Verlag},\ \bibinfo {address}
  {Berlin},\ \bibinfo {year} {2008})%
  \bibAnnoteFile{NoStop}{SchlosshauerText}%
\bibitem{JoosText}%
  \BibitemOpen
  \bibfield{author}{%
  \bibinfo {author} {\bibfnamefont{E.}~\bibnamefont{Joos}}, \bibinfo {author}
  {\bibfnamefont{H.~D.}\ \bibnamefont{Zeh}}, \bibinfo {author}
  {\bibfnamefont{C.}~\bibnamefont{Kiefer}}, \bibinfo {author}
  {\bibfnamefont{D.}~\bibnamefont{Giulini}}, \bibinfo {author}
  {\bibfnamefont{J.}~\bibnamefont{Kupsch}},\ and\ \bibinfo {author}
  {\bibfnamefont{I.-O.}\ \bibnamefont{Stamatescu}},\ }%
  \emph{\bibinfo {title} {Decoherence and the Appearance of the Classical World
  in Quantum Theory}}\ (\bibinfo {publisher} {Springer-Verlag},\ \bibinfo
  {address} {Berlin},\ \bibinfo {year} {2003})%
  \bibAnnoteFile{NoStop}{JoosText}%
\bibitem{Zurek2003}%
  \BibitemOpen
  \bibfield{author}{%
  \bibinfo {author} {\bibfnamefont{W.~H.}\ \bibnamefont{Zurek}},\ }%
  \bibfield{journal}{%
  \Doi{10.1103/RevModPhys.75.715}{\bibinfo {journal} {Rev. Mod. Phys.}}\ }%
  \textbf{\bibinfo {volume} {75}},\ \bibinfo {pages} {715} (\bibinfo {month}
  {May}\ \bibinfo {year} {2003})%
  \bibAnnoteFile{NoStop}{Zurek2003}%
\bibitem{Zurek2009}%
  \BibitemOpen
  \bibfield{author}{%
  \bibinfo {author} {\bibfnamefont{W.~H.}\ \bibnamefont{Zurek}},\ }%
  \bibfield{journal}{%
  \bibinfo {journal} {Nature Physics}\ }%
  \textbf{\bibinfo {volume} {5}},\ \bibinfo {pages} {181} (\bibinfo {year}
  {2009})%
  \bibAnnoteFile{NoStop}{Zurek2009}%
\bibitem{Zurek2000}%
  \BibitemOpen
  \bibfield{author}{%
  \bibinfo {author} {\bibfnamefont{W.~H.}\ \bibnamefont{Zurek}},\ }%
  \bibfield{journal}{%
  \Doi{10.1002/1521-3889(200011)9:11/12<855::AID-ANDP855>3.0.CO}{\bibinfo
  {journal} {Ann. Phys. (Leipzig)}}\ }%
  \textbf{\bibinfo {volume} {9}},\ \bibinfo {pages} {855} (\bibinfo {month}
  {Sept}\ \bibinfo {year} {2000})%
  \bibAnnoteFile{NoStop}{Zurek2000}%
\bibitem{Ollivier2004}%
  \BibitemOpen
  \bibfield{author}{%
  \bibinfo {author} {\bibfnamefont{H.}~\bibnamefont{Ollivier}}, \bibinfo
  {author} {\bibfnamefont{D.}~\bibnamefont{Poulin}},\ and\ \bibinfo {author}
  {\bibfnamefont{W.~H.}\ \bibnamefont{Zurek}},\ }%
  \bibfield{journal}{%
  \Doi{10.1103/PhysRevLett.93.220401}{\bibinfo {journal} {Phys. Rev. Lett.}}\
  }%
  \textbf{\bibinfo {volume} {93}},\ \bibinfo {pages} {220401} (\bibinfo {month}
  {Nov}\ \bibinfo {year} {2004})%
  \bibAnnoteFile{NoStop}{Ollivier2004}%
\bibitem{Blume-Kohout2005}%
  \BibitemOpen
  \bibfield{author}{%
  \bibinfo {author} {\bibfnamefont{R.}~\bibnamefont{Blume-Kohout}}\ and\
  \bibinfo {author} {\bibfnamefont{W.~H.}\ \bibnamefont{Zurek}},\ }%
  \bibfield{journal}{%
  \Doi{10.1007/s10701-005-7352-5}{\bibinfo {journal} {Found. Phys.}}\ }%
  \textbf{\bibinfo {volume} {35}},\ \bibinfo {pages} {1857} (\bibinfo {year}
  {2005})%
  \bibAnnoteFile{NoStop}{Blume-Kohout2005}%
\bibitem{Zwolak2009}%
  \BibitemOpen
  \bibfield{author}{%
  \bibinfo {author} {\bibfnamefont{M.}~\bibnamefont{Zwolak}}, \bibinfo {author}
  {\bibfnamefont{H.~T.}\ \bibnamefont{Quan}},\ and\ \bibinfo {author}
  {\bibfnamefont{W.~H.}\ \bibnamefont{Zurek}},\ }%
  \bibfield{journal}{%
  \Doi{10.1103/PhysRevLett.103.110402}{\bibinfo {journal} {Phys. Rev. Lett.}}\
  }%
  \textbf{\bibinfo {volume} {103}},\ \bibinfo {eid} {110402} (\bibinfo {year}
  {2009})%
  \bibAnnoteFile{NoStop}{Zwolak2009}%
\bibitem{Zwolak2009a}%
  \BibitemOpen
  \bibfield{author}{%
  \bibinfo {author} {\bibfnamefont{M.}~\bibnamefont{Zwolak}}, \bibinfo {author}
  {\bibfnamefont{H.}~\bibnamefont{Quan}},\ and\ \bibinfo {author}
  {\bibfnamefont{W.~H.}\ \bibnamefont{Zurek}},\ }%
  \bibfield{journal}{%
  \Doi{10.1103/PhysRevA.81.062110}{\bibinfo {journal} {Phys. Rev. A}}\ }%
  \textbf{\bibinfo {volume} {81}},\ \bibinfo {pages} {062110} (\bibinfo {year}
  {2010})%
  \bibAnnoteFile{NoStop}{Zwolak2009a}%
\bibitem{Blume-Kohout2008}%
  \BibitemOpen
  \bibfield{author}{%
  \bibinfo {author} {\bibfnamefont{R.}~\bibnamefont{Blume-Kohout}}\ and\
  \bibinfo {author} {\bibfnamefont{W.~H.}\ \bibnamefont{Zurek}},\ }%
  \bibfield{journal}{%
  \Doi{10.1103/PhysRevLett.101.240405}{\bibinfo {journal} {Phys. Rev. Lett.}}\
  }%
  \textbf{\bibinfo {volume} {101}},\ \bibinfo {eid} {240405} (\bibinfo {year}
  {2008})%
  \bibAnnoteFile{NoStop}{Blume-Kohout2008}%
\bibitem{Paz2009}%
  \BibitemOpen
  \bibfield{author}{%
  \bibinfo {author} {\bibfnamefont{J.~P.}\ \bibnamefont{Paz}}\ and\ \bibinfo
  {author} {\bibfnamefont{A.~J.}\ \bibnamefont{Roncaglia}},\ }%
  \bibfield{journal}{%
  \Doi{10.1103/PhysRevA.80.042111}{\bibinfo {journal} {Physical Review A}}\ }%
  \textbf{\bibinfo {volume} {80}},\ \bibinfo {pages} {042111} (\bibinfo {year}
  {2009})%
  \bibAnnoteFile{NoStop}{Paz2009}%
\bibitem{Joos1985}%
  \BibitemOpen
  \bibfield{author}{%
  \bibinfo {author} {\bibfnamefont{E.}~\bibnamefont{Joos}}\ and\ \bibinfo
  {author} {\bibfnamefont{H.~D.}\ \bibnamefont{Zeh}},\ }%
  \bibfield{journal}{%
  \Doi{10.1007/BF01725541}{\bibinfo {journal} {Zeitschrift f\"{u}r Physik B
  Condensed Matter}}\ }%
  \textbf{\bibinfo {volume} {59}},\ \bibinfo {pages} {223} (\bibinfo {year}
  {1985})%
  \bibAnnoteFile{NoStop}{Joos1985}%
\bibitem{Diosi1995}%
  \BibitemOpen
  \bibfield{author}{%
  \bibinfo {author} {\bibfnamefont{L.}~\bibnamefont{Diosi}},\ }%
  \bibfield{journal}{%
  \Doi{10.1209/0295-5075/30/2/001}{\bibinfo {journal} {Europhys. Lett.}}\ }%
  \textbf{\bibinfo {volume} {30}},\ \bibinfo {pages} {63} (\bibinfo {year}
  {1995})%
  \bibAnnoteFile{NoStop}{Diosi1995}%
\bibitem{Gallis1990}%
  \BibitemOpen
  \bibfield{author}{%
  \bibinfo {author} {\bibfnamefont{M.~R.}\ \bibnamefont{Gallis}}\ and\ \bibinfo
  {author} {\bibfnamefont{G.~N.}\ \bibnamefont{Fleming}},\ }%
  \bibfield{journal}{%
  \Doi{10.1103/PhysRevA.42.38}{\bibinfo {journal} {Phys. Rev. A}}\ }%
  \textbf{\bibinfo {volume} {42}},\ \bibinfo {pages} {38} (\bibinfo {month}
  {Jul}\ \bibinfo {year} {1990})%
  \bibAnnoteFile{NoStop}{Gallis1990}%
\bibitem{Hornberger2003}%
  \BibitemOpen
  \bibfield{author}{%
  \bibinfo {author} {\bibfnamefont{K.}~\bibnamefont{Hornberger}}\ and\ \bibinfo
  {author} {\bibfnamefont{J.~E.}\ \bibnamefont{Sipe}},\ }%
  \bibfield{journal}{%
  \Doi{10.1103/PhysRevA.68.012105}{\bibinfo {journal} {Phys. Rev. A}}\ }%
  \textbf{\bibinfo {volume} {68}},\ \bibinfo {pages} {012105} (\bibinfo {month}
  {Jul}\ \bibinfo {year} {2003})%
  \bibAnnoteFile{NoStop}{Hornberger2003}%
\bibitem{Hornberger2006}%
  \BibitemOpen
  \bibfield{author}{%
  \bibinfo {author} {\bibfnamefont{K.}~\bibnamefont{Hornberger}},\ }%
  \bibfield{journal}{%
  \Doi{10.1103/PhysRevLett.97.060601}{\bibinfo {journal} {Phys. Rev. Lett.}}\
  }%
  \textbf{\bibinfo {volume} {97}},\ \bibinfo {pages} {060601} (\bibinfo {year}
  {2006})%
  \bibAnnoteFile{NoStop}{Hornberger2006}%
\bibitem{Kokorowski2001}%
  \BibitemOpen
  \bibfield{author}{%
  \bibinfo {author} {\bibfnamefont{D.~A.}\ \bibnamefont{Kokorowski}}, \bibinfo
  {author} {\bibfnamefont{A.~D.}\ \bibnamefont{Cronin}}, \bibinfo {author}
  {\bibfnamefont{T.~D.}\ \bibnamefont{Roberts}},\ and\ \bibinfo {author}
  {\bibfnamefont{D.~E.}\ \bibnamefont{Pritchard}},\ }%
  \bibfield{journal}{%
  \Doi{10.1103/PhysRevLett.86.2191}{\bibinfo {journal} {Phys. Rev. Lett.}}\ }%
  \textbf{\bibinfo {volume} {86}},\ \bibinfo {pages} {2191} (\bibinfo {month}
  {Mar}\ \bibinfo {year} {2001})%
  \bibAnnoteFile{NoStop}{Kokorowski2001}%
\bibitem{Uys2005}%
  \BibitemOpen
  \bibfield{author}{%
  \bibinfo {author} {\bibfnamefont{H.}~\bibnamefont{Uys}}, \bibinfo {author}
  {\bibfnamefont{J.~D.}\ \bibnamefont{Perreault}},\ and\ \bibinfo {author}
  {\bibfnamefont{A.~D.}\ \bibnamefont{Cronin}},\ }%
  \bibfield{journal}{%
  \Doi{10.1103/PhysRevLett.95.150403}{\bibinfo {journal} {Phys. Rev. Lett.}}\
  }%
  \textbf{\bibinfo {volume} {95}},\ \bibinfo {pages} {150403} (\bibinfo {month}
  {Oct}\ \bibinfo {year} {2005})%
  \bibAnnoteFile{NoStop}{Uys2005}%
\bibitem{JacksonEMcrosssection}%
  \BibitemOpen
  \bibfield{author}{%
  \bibinfo {author} {\bibfnamefont{J.~D.}\ \bibnamefont{Jackson}},\ }%
  \enquote{\bibinfo {title} {Classical electrodynamics},}\ \ (\bibinfo
  {publisher} {John Wiley},\ \bibinfo {address} {New York},\ \bibinfo {year}
  {1999})\ p.\ \bibinfo {pages} {459},\ \bibinfo {edition} {3rd}\ ed.%
  \bibAnnoteFile{Stop}{JacksonEMcrosssection}%
\bibitem{Note1}%
  \BibitemOpen
  \bibinfo {note} {The decoherence rate does not increase for arbitrarily large
  $\Delta x$. Rather, this expression is only valid in the $\Delta x \ll
  \lambda $ limit we are considering here. For $\Delta x \gg \lambda $, the
  decoherence rate saturates \begin {align*} \protect \frac {1}{\protect
  \mathaccentV {tilde}07E{\tau }_D} = \protect \mathaccentV
  {tilde}07E{C}_\Gamma \protect \frac {I \protect \mathaccentV {tilde}07E{a}^6
  k_B^3 T^3}{c^4 \hbar ^4} \end {align*} with $\protect \mathaccentV
  {tilde}07E{C}_\Gamma = 57600 \protect \tmspace +\thinmuskip {.1667em} \zeta
  (7)/\pi ^3 \approx 1873$, in agreement with \cite {Gallis1990}. In the
  intermediate region, $\Delta x \sim \lambda $, $\tau _D$ has a complicated
  dependence on both $\Delta x$ and $\theta $. The results of the present work
  are valid for \protect \emph {all} $\Delta x$ so long as the correct $\tau
  _D$ is used.}%
  \bibAnnoteFile{Stop}{Note1}%
\bibitem{Blume-Kohout2006}%
  \BibitemOpen
  \bibfield{author}{%
  \bibinfo {author} {\bibfnamefont{R.}~\bibnamefont{Blume-Kohout}}\ and\
  \bibinfo {author} {\bibfnamefont{W.~H.}\ \bibnamefont{Zurek}},\ }%
  \bibfield{journal}{%
  \Doi{10.1103/PhysRevA.73.062310}{\bibinfo {journal} {Phys. Rev. A}}\ }%
  \textbf{\bibinfo {volume} {73}},\ \bibinfo {pages} {062310} (\bibinfo {month}
  {Jun}\ \bibinfo {year} {2006})%
  \bibAnnoteFile{NoStop}{Blume-Kohout2006}%
\bibitem{Bennett2006}%
  \BibitemOpen
  \bibfield{author}{%
  \bibinfo {author} {\bibfnamefont{C.~H.}\ \bibnamefont{Bennett}},\ }%
  in\ \emph{\bibinfo {booktitle} {Quantum Computing: Back Action 2006}},\
  \bibinfo {editor} {edited by\ \bibinfo {editor}
  {\bibfnamefont{D.}~\bibnamefont{Goswami}}}\ (\bibinfo {publisher} {AIP},\
  \bibinfo {address} {New York},\ \bibinfo {year} {2006})\ pp.\ \bibinfo
  {pages} {11--17}%
  \bibAnnoteFile{NoStop}{Bennett2006}%
\end{thebibliography}%

%%%%%%%%%%%%%%%%%%%%%%%%%%%%%
\end{document}